# Region of Interest Identification for Brain Tumors in Magnetic Resonance Images

Fateme Mostafaie*, Reihaneh Teimouri*, Zahra Nabizadeh, Nader Karimi, Shadrokh Samavi
Department of Electrical and Computer Engineering,
Isfahan University of Technology,
Isfahan, 84156-83111 Iran

*Abstract*— Glioma is a common type of brain tumor, and accurate detection of it plays a vital role in the diagnosis and treatment process. Despite advances in medical image analyzing, accurate tumor segmentation in brain magnetic resonance (MR) images remains a challenge due to variations in tumor texture, position, and shape. In this paper, we propose a fast, automated method, with light computational complexity, to find the smallest bounding box around the tumor region. This region-of-interest can be used as a preprocessing step in training networks for subregion tumor segmentation. By adopting the outputs of this algorithm, redundant information is removed; hence the network can focus on learning notable features related to subregions' classes. The proposed method has six main stages, in which the brain segmentation is the most vital step. Expectation-maximization (EM) and K-means algorithms are used for brain segmentation. The proposed method is evaluated on the BraTS 2015 dataset, and the average gained DICE score is 0.73, which is an acceptable result for this application.

## I. Introduction

A brain tumor is formed by the uncontrolled growth of abnormal cells in the brain. Some brain tumors are noncancerous (benign), and the others are cancerous (malignant). Benign brain tumors do not contain cancer cells and have a homogeneous structure, while malignant brain tumors contain cancer cells and have a heterogeneous structure. Brain tumors can be categorized as primary or secondary. Primary brain tumors start in the brain and stay there. Secondary brain tumors, also known as a metastatic brain tumor, are more common popular and occur when cancer cells spread to the brain from another organ, such as Lung, breast, kidney, colon, and skin. The most popular primary brain tumor is the glioma type. They grow from a type of brain cell called a glial cell. Gliomas can be categorized into two basic grades: low-grade gliomas (LGG) that the tumor cells look more slowly dividing under the microscope and high-grade gliomas (HGG) that the cells look more aggressive under the microscope [1].

Automatic segmentation of brain tumors and subregions can play an active role in better diagnosis, surgical planning, and treatment process of brain tumors. However, different methods proposed in recent years, but it is still challenging because size, shape, and location of brain tumors vary among patients, and also the boundaries between adjacent regions are often ambiguous [2].

With the development of medical imaging, brain tumors can be imaged by neural imaging modalities, mainly computed tomography (CT) and magnetic resonance imaging (MRI). In the Brain Tumor Image Segmentation (BraTS) 2015 dataset, all patient's samples are 3D MRI. There are four modalities for each patient: T1-weighted, contrast-enhanced T1-weighted (T1c), T2-weighted and fluid-attenuated inversion recovery (FLAIR). Each sequence can provide complementary information for segmenting different subregions of a tumor, which are defined as follows: edema, non-enhancing (solid) core, necrotic (or fluid-filled) core, and non-enhancing core. For instance, T2 and FLAIR highlight the whole tumor region, or T1 and T1c highlight the tumor's core area [3].

There are different methods for segmentation. Segmentation approaches could be divided into two groups, (i) traditional image segmentation and (ii) segmentation based on neural networks. For traditional image segmentation, strategies like histogram-based image segmentation [4], edge detection based [5], texture features based [6], and Super-voxel segmentation [7] could be mentioned. On the other hand, there are many approaches based on neural network algorithms from 2D neural networks like [8] to 3D neural networks like [2]. Traditional image segmentation approaches are tried to extract features like edge, texture, color, etc. that distinct objects from their background. In other words, features that are different between backgrounds and objects are extracted. In neural network approaches, networks try to extract features and then up-sample and build segmented images again. Most of the time, they may be accurate as networks extract features, and they are not handcrafted.

In [9], a bounding box was extracted based on T2 and FLAIR due to decrease computational time and memory. In this work, more relevant patches that contain tumors are selected, to avoid using images that have missing parts. In [2], three networks have been proposed to hierarchically segment whole tumor, tumor core, and enhancing tumor core sequentially, so the model converts the problem into three smaller binary segmentation problems. The first 3D neural network, WNet, was employed to find a bounding box around the tumor in 3D MRI, in order to sequentially segment sub-tumors. The output of WNet is used as a mask to crop the whole tumor. In [10], to reduce network complexity and memory consumption, a 2.5D network is used. Similar to [2], in this method, a network was designed to extract the whole brain tumor area to find sub-tumors easier.

As mentioned above, the primary step before finding the subregions of tumors in brain tumor segmentation is to find the whole tumor. Existing methods use heavy computational tools for extracting the entire tumor region like training networks. In this paper, we propose a novel approach that

---

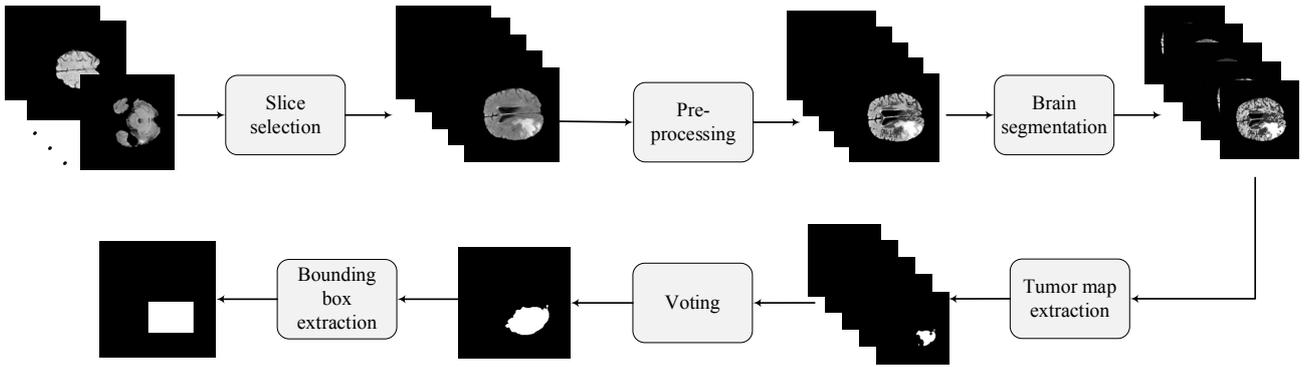

Fig. 1. Workflow of the proposed method

gives the smallest bounding box containing the region of the tumor without relying on computationally expensive algorithms. The cropping algorithm includes six main blocks, including slice selection, pre-processing, brain segmentation, tumor map extraction, voting, and bounding box extraction.

The rest of the paper is organized as follows. Section II introduces the proposed method in detail. Experimental results are presented in Section III. Finally, Section IV gives the conclusion.

## II. PROPOSED METHOD

The workflow for finding the smallest bounding box containing the tumor region of each MRI case involves the six main steps, which are shown in Fig. 1. In this work, processing of slices is chosen instead of the processing 3D data so, before going through the principal blocks, 2D MRI slices are extracted from the 3D volume, and among four modalities, FLAIR one is used as the input of main blocks. At first, in the slice selection block, 6 slices are chosen as a representation of all 155 slices. Secondly, in the pre-processing block, the contrast of each slice is changed to make it appropriate for the segmentation block. Then, the segmentation algorithm is used to segment each of the enhanced slices. After that, the tumor map corresponding to each segmented slice is generated in the tumor map extraction block. In this step, six tumor maps are selected that voting between them can make the final tumor map. In the end, the smallest rectangle around the tumor region is extracted from the final tumor map. In the following, each block is explained in more detail.

### A. Slice selection

The slice-selection block aims to reduce computational time by decreasing the number of input slices for the next blocks. The tumor region appears in one slice and becomes more substantial in the following slices. By considering this fact, we can choose some representative slices instead of working on all 155 slices. The representative slices should be ones that the tumor region has the largest size in them. With satisfying this condition, we can make sure that the final extracted bounding box contains the tumor region in all slices. For finding the appropriate representative slices, in the first step, all slices from 1 to 31 and from 119 to 155 are omitted because they do not contain any region of the brain. Then by analyzing the ground truth of all patients in the dataset, six slices are chosen where the tumor regions are the largest, and these slices are common among all patients in the dataset. For all patients, for each slice, the number of tumor pixels is computed. Slices with the largest tumor regions are selected. Representative slices are numbers 50, 66, 87, 89, 92, and 110. These slices are fed into the next block as an input. These slices are selected by statistical analysis of the train data and used for test data.

### B. Pre-processing

Because of different acquisition techniques and systems, the MR images, when extracted from volumetric data, have artifacts [10]. In the pre-processing block, each slice is normalized with respect to its minimum and maximum pixel values to make input slices more suitable for the next step. Next, the tumor region intensity is increased while healthy region intensity is decreased. In Fig. 2, the result of the pre-processing block is shown. More information about each sub-block is provided below.

*1) Slice normalization:* In this sub-block, a gray-level normalization is performed for each input slice. For an input image $I$, the normalized image is derived based on (1).

$$I_n = (I - min)/(max - min) \qquad (1)$$

where $I_n$ is the normalized image, $min$ and $max$ are the minimum and maximum intensities of the input image. Normalization causes processing all slices similarly, considering the fact that different slices may have a different grayscale range of values. MRI of different patients may have been captured by dissimilar MRI machines with different Tesla values.

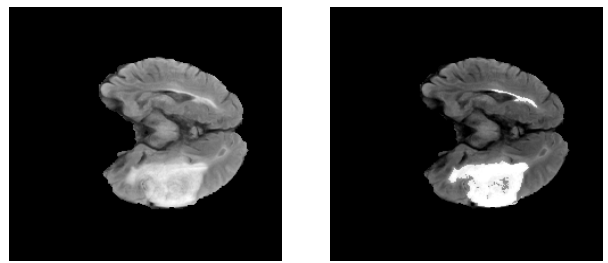

Fig. 2. Input slice (left) and output slice (right) of pre-processing block

*2) Tumor contrast enhancement:* In tumor detection problem, it is crucial to improve tumor region contrast as compared to healthy pixels. Thus the performance of methods is related to good contrast stretching, which makes the tumor region more visible [11]. Therefore, finding the pixels which are in the tumor region is the challenge. For this purpose, two features are used. The first feature is the intensity of each pixel, and the second feature is the weight of the pixel. The value of each feature depends on the location of the pixel being inside or outside of the tumor region. For checking the intensity, in this sub-block, at first, a slice-based threshold is found. Due to the black area around each image, for choosing the best threshold value, the average intensity of brain region pixels is used, and the effect of the black area is omitted. After finding the threshold, the pixels which are more likely to be the tumor pixels should be found. For doing this, a slice-based location atlas is utilized [12]. The location atlas is computed for each slice and is derived from the summation of the binary ground truth of all patients. For six representative slices mentioned before, six atlases are generated. Each location atlas is computed using (2).

$$l_n = \sum_{i,j}^{W,H} \sum_{k}^{P} p_{k,n}(i,j) \qquad (2)$$

Where n∈ {50,66,87,98,92,110} is the atlas number, P is the number of patients, W and H are the width and height of slices. Also $p_{k,n}(i,j)$ is defined as (3).

$$p_{k,n}(i,j) = \begin{cases} 1 & if\ v_{k,n}(i,j) \neq 0 \\ 0 & otherwise \end{cases} \qquad (3)$$

Where $v_{k,n}(i,j)$ is the value of pixel (i,j) in slice $n$ for patient $k$ in the ground truth. Pixels that their values are more than zero are the more likely tumor pixels. Therefore, the contrast of selected tumor pixels should be improved. If intensities of these pixels are more than the calculated threshold, their intensities increase, and if their values are less, their intensities are decreased.

*C. Brain segmentation*

The essential block in the proposed workflow is the brain segmentation because it detects the tumor region in the input slice. Several methods are implemented recently to segment the brain region. In this work, we use two famous EM and K-means algorithms. Each input slice is segmented into five classes by both algorithms. The results in Fig. 3, show that by using the uncertainty degree of assignment, the output of EM would be better than K-means. In the following, both methods are described in more detail.

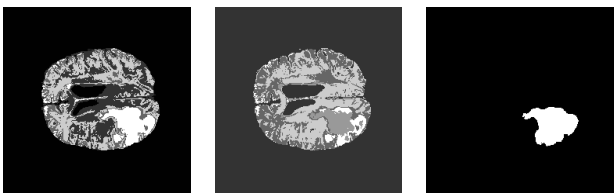

Fig. 3. EM segmentation result (left), K-means segmentation result (middle), ground truth(right)

*1) EM algorithm:* The EM algorithm is used to find a maximum-likelihood estimation for model variables when a dataset is incomplete and has missing data, or latent variables (variables that are not directly observable and are inferred from the values of the other observed variables). This approach stems from the Gaussian Mixture Model (GMM). The EM Algorithm contains two steps to improve a parameter's estimation. These iterations continue until algorithm convergence. This algorithm needs a random start variable to run from the local maxima that are not close to the (optimal) global maxima. The first step, E-step, estimates the missing variables in the dataset by calculating the probability of given observed data. Next, in the M-Step, whole data is used to update the parameters and keeping the hidden values fixed [13]. The advantage of the EM algorithm over other clustering algorithms, such as K-means, is that EM uses soft assignment against hard assignment. It means that in soft assignment every pixel belongs to groups with the uncertainty degree. Whereas in the hard assignment, a point belongs to a group with complete certainty. In our proposed approach, the intensity of each pixel is input value for EM, and the Gaussian mixture model represents five classes of pixel intensities in each slice.

*2) K-means algorithm:* K-means algorithm is one of the most straightforward unsupervised machine learning approaches. First, the K-means algorithm based on the number of clusters selects the centroids. Every centroid represents one cluster. Then the algorithm allocates every pixel to the nearest cluster to minimize the distance between each pixel and its centroid. In the next step, K-means tries to find the best centroids to further reduce the distance between the new centroid and elements of the cluster. This iterative algorithm continues until converging to the minimum distance by optimizing the position of centroids in every iteration [14]. The number of clusters is 5 in our method.

*D. Tumor map extraction*

In this block, the tumor region is extracted from the segmented slice. From 5 classes, regions labeled as 4 or 5 are tumor regions and should be extracted. This process is done in 4 steps:

(1) The biggest connected component region, which labeled as 5 is extracted.

(2) If the extracted region has a suitable area (not too big or too small), then all pixels with a specific radius around the region's center is considered as a tumor.

(3) Else, step 1, and step 2 are repeated for the region, which labeled as 4.

(4) If no region is extracted, the tumor map is black, and no tumor is detected in that slice.

It is crucial to consider the condition of the second step because in some bright slices, the whole brain region detected as class 5 or 4, and it causes the entire brain area detected as the tumor.

*E. Voting*

Combining all six provided tumor maps and finding the final tumor map is a very challenging task., a new voting approach is used to combine all six maps efficiently. In this approach, each slice is divided into four parts. Then the number of slices that detect the tumor region in each part is calculated. Those parts that have more than one votes are chosen, and the union of detected tumor regions in them is extracted as the final tumor map. In Fig. 4, six tumor maps, for slice numbers 50, 66, 87, 89, 92, 110, of one patient are shown. Parts 3 and 4 get more than one vote; therefore, these two parts are chosen, and the union of detected regions on them is extracted as the final tumor map.

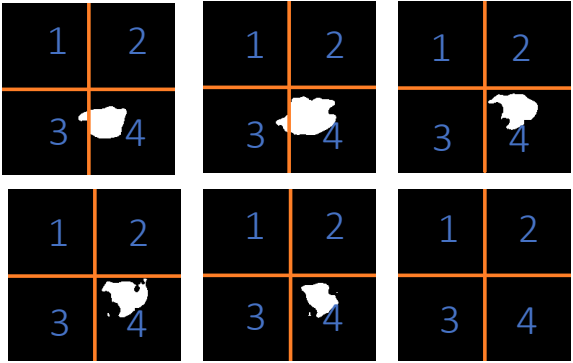

Fig. 4. The voting operation on six tumor maps.

*F. Bounding box extraction*

At the final step, the smallest rectangle that contains the tumor region is extracted. This bounding box is the final output that can be used as a mask for cropping tumor regions in the pre-processing step of sub-region segmentation methods.

## III. EXPERIMENTS AND RESULTS

*A. Dataset and setting*

We used BraTS 2015 dataset to evaluate our method. This database contains image volumes from 210 patients for HGG and 75 patients for LGG. The volumes are skull stripped and registered. Each volume contains multiple slices of MR images. As mentioned before, the image dataset consists of four MRI types, including T1, T1c, FLAIR, and T2. Also, the dataset includes the ground truth, which is composed of five labeled areas. Area 1 is for necrosis, label 2 for edema, label 3 for non-enhancing tumor, label 4 for enhancing tumor, and label 0 is for everything else. We need a ground truth with just two labels that show the tumor and non-tumor areas. Hence, we consider a new ground truth that was made based on (4), where $GT(i,j)$ shows the corresponding label for a pixel, which has $(i,j)$ coordinates.

$$New\_GT(i,j) = \begin{cases} 1 & if\ GT(i,j) \neq 0 \\ 0 & otherwise \end{cases} \quad (4)$$

Then a cumulative ground truth for a specific patient was obtained from adding all 155 slices using (5).

$$Cumulative\_GT(i,j) = \begin{cases} 1 & if\ \sum_{k=1}^{155} New\_GT_k(i,j) > 0 \\ 0 & otherwise \end{cases} \quad (5)$$

In the end, the final ground truth was produced by considering the smallest rectangle that covers all parts of the tumor in the Cumulative_GT.

*B. Evaluation metrics*

In this proposed approach, the Dice score is used for measuring our method's performance. It ranges from 0 to 1, that 1 shows the highest similarity between the predicted crop region and the ground truth. If the set A is considered as pixels of the smallest rectangle containing Cumulative_GT and the set B is considered as pixels of the generated rectangle from our proposed method, then the dice score is computed by (6).

$$Dice\ score(A,B) = \frac{2 \times |A \cap B|}{|A \cup B|} \quad (6)$$

*C. Experiments and results*

The proposed method is implemented by Matlab 2019Ra. EM and K-mean algorithms are two clustering algorithms that are employed for the segmentation block, which produce two different final results. In Table 1, the results of the two segmentation algorithms applied to each group are shown. It shows that the EM algorithm is more successful than K-means.

Table 1. Dice score results from EM and K-means segmentation algorithms

| Method | Dice (HGG) | Dice (LGG) |
|---|---|---|
| **EM** | 0.75 | 0.69 |
| **K-means** | 0.55 | 0.50 |

## IV. CONCLUSION

In this paper, we proposed an approach for finding region-of-interest in brain MR images. The region-of-interest is formed by cropping the smallest patch that contains the whole tumor region. Since a large number of slices do not include any tumor region, it is essential to omit the healthy parts of the brain to balance the network's input data. Using this cropping method as a pre-processing step for tumor segmentation networks can help improve their performance. In other research approaches, the whole tumor area is detected with complex networks to omit unwanted brain parts and find sub-tumors with other networks, whereas, in this method, the best patch is chosen by the help of the clustering algorithms on only six slices instead of 155 slices. Dice score of these methods on both HGG and LGG images is 0.73, and the result can be used with the safe margin to confirm that the whole tumor is in the selected crop. Therefore, in this way, not only no part of the tumor will be lost, but the input data will be balanced, and unwanted parts will be completely ignored with a low computational method.